\documentclass[prd,preprint]{revtex4}

\usepackage{xcolor}

	\usepackage{amsmath}
	\usepackage{amsfonts}
  	\usepackage{amssymb}
	\usepackage{makeidx}
	\usepackage{amsfonts}
	\usepackage[ansinew]{}
	\usepackage[usenames,dvipsnames]{pstricks}
	\usepackage{epsfig}

    \usepackage{graphicx}

	\usepackage{float}

\newcommand{\be}{\begin{equation}}
\newcommand{\ee}{\end{equation}}
\newcommand{\bea}{\begin{eqnarray}}
\newcommand{\eea}{\end{eqnarray}}
\newcommand{\ba}{\begin{align}}
\newcommand{\ea}{\end{align}}

\newcommand\SPD{\mathrel{\stackrel{\makebox[0pt]{\mbox{\normalfont\tiny (3)}}}{\Delta}}}

\newcommand{\dd}{{\mathrm{d}}}

\newcommand\R{\mathcal{R}}	
\makeindex

\begin{document}

\title{Are primordial black holes produced by entropy perturbations in single field inflationary models?}
\author{Sergio Andr\'es Vallejo-Pe\~na$^{2,3}$}
\author{Antonio Enea Romano$^{1,2,3,4}$}

\affiliation{
${}^{1}$Theoretical Physics Department, CERN, CH-1211 Geneva 23, Switzerland\\
${}^{2}$ICRANet, Piazza della Repubblica 10, I--65122 Pescara, Italy \\
${}^3$Instituto de Fisica, Universidad de Antioquia, A.A.1226, Medellin, Colombia
\\
${}^4$Department of Physics \& Astronomy, Bishop’s University
2600 College Street, Sherbrooke, Qu´ebec, Canada J1M 1Z7
}


\begin{abstract}
We show that in  single field inflationary models the super-horizon evolution of  curvature perturbations on comoving slices $\R$, which can cause the production of primordial black holes (PBH), is not due to entropy perturbations, but to the background evolution effect on the conversion between entropy and curvature perturbations. 
We derive a  general  relation between  the time derivative of comoving curvature perturbations and entropy perturbations, in terms of a conversion factor depending on the background evolution. Contrary to previous results derived in the uniform density gauge assuming the gradient term can be neglected on super-horizon scales, the relation is  valid on any scale for any minimally coupled single scalar field model, also on sub-horizon scales where gradient terms are large. 

We apply it to the case of quasi-inflection inflation, showing that while entropy perturbations are decreasing, $\R$ can grow on super-horizon scales, due to a large increase of the conversion factor. This happens in the time interval during which a sufficiently fast decrease of the equation of state $w$ transforms into a growing mode what in slow-roll models would be a decaying mode. The same mechanism also explains the super-horizon evolution of $\R$ in globally adiabatic systems, for which entropy perturbations vanish  on any scale, such as ultra slow roll inflation and its generalizations.
\end{abstract}


\maketitle

\section{Introduction}
According to the standard cosmological model primordial curvature perturbations provided the seeds from which large scale structure has formed.
When these perturbations are sufficiently large \cite{Sasaki:2018dmp} primordial black holes (PBH) could be formed, with a series of important observational consequences \cite{Sasaki:2016jop,Josan:2009qn,Carr:2009jm,Belotsky:2014kca,Belotsky:2018wph}.
In this paper we investigate what are the general  conditions for the super-horizon evolution of comoving curvature perturbations $\R$ in single field models.

In  slow-roll inflationary models $\R$ is conserved on super-horizon scales \cite{Wands:2000dp}. However there are other single  field models with  super-horizon evolution of $\R$, such as globally adiabatic models \cite{Romano:2016gop} or inflation with a quasi-inflection point in the potential \cite{Ezquiaga:2018gbw}. 
The super-horizon growth of $\R$ has profound implications because it can give rise to PBH production  \cite{Garcia-Bellido:2017mdw}, with different important observable effects.

In the case of globally adiabatic models \cite{Romano:2016gop}, which include ultra slow roll inflation \cite{Tsamis:2003px,Kinney:2005vj} for example,
it was already shown that the cause of the super-horizon growth of $\R$ are not entropy perturbations, since those modes are adiabatic on any scale.
For quasi-inflection inflation instead it has been argued \cite{Ezquiaga:2018gbw,Leach:2000yw} that the  super-horizon growth of $\R$ is due to entropy perturbations. 
Nevertheless, in agreement with the general analysis given in \cite{Romano:2016gop}, it has also been  \cite{Rasanen:2018fom,Biagetti:2018pjj,Cicoli:2018asa,Ozsoy:2018flq} shown in different scenarios that the super-horizon growth of $\R$ can be explained uniquely in terms of the evolution of the background, so it is important to clarify if non adiabatic perturbations actually play any important role.  

In order to better understand the mechanism producing this phenomena we analyze the general relation between curvature and entropy perturbation in single field models, showing that the quantity which plays the most important role is not the entropy perturbation but the equation of state $w=P/\rho$, whose fast time variation can induce a super-horizon evolution of what in slow-roll models would be a decaying mode.   
As an application we show that similarly to what happens for globally adiabatic models, also for quasi-inflection models the super-horizon growth is not due to entropy perturbation, but to the background evolution. 
\section{Single scalar field models}
The Lagrangian for single scalar fields models minimally coupled to gravity is
\begin{align}
    S &= \int \dd^4 x \sqrt{-g} \left[ \frac{1}{2} M_p^2 R - \frac{1}{2} g^{\mu \nu} \partial_\mu \phi \partial_\nu \phi - V(\phi) \right] \, , \label{Action} 
\end{align}
where 
$R$ is the Ricci scalar, $g^{\mu \nu}$ is the FLRW metric 
\be
\dd s^2 =-\dd t^2+a(t)^2\delta_{ij}\dd x^i\dd x^j \, , 
\ee
and we are using  a system of units in which $c=\hslash=1$ and $M_p= (8\pi G)^{-\frac{1}{2}}$ is the reduced Planck mass. 
The variation of the action with respect to the metric and the scalar field gives 
\begin{align}
 H^2 &= \frac{1}{3 M_p^2} \left( \frac{\dot{\phi}^2}{2} + V(\phi) \right) \, , \label{eqH} \\
 \ddot{\phi} &+ 3 H \dot{\phi} + V_{\phi} = 0 \, , \label{eqphi}
\end{align}
where a dot denotes a derivative with respect to cosmic time $t$, $H=\dot{a}/a$ is the Hubble parameter, and $V_{\phi}=\partial_{\phi} V(\phi)$.
The background energy momentum tensor of the scalar field is the same of a perfect fluid with energy density $\rho=\frac{1}{2}\dot{\phi}^2+V$ and pressure $P=\frac{1}{2}\dot{\phi}^2-V$.
We define the slow-roll parameters as 
\begin{align}
    \epsilon &\equiv - \frac{\dot{{H}}}{H^2} \, , \label{epsilon} \\
    \eta &\equiv \frac{\dot{\epsilon}}{\epsilon H} \, . \label{eta}
\end{align}

In the next section we will study some properties of the super-horizon behavior of $\R$ for this class  of models without making any specific choice for the potential, reaching some general conclusions about the conditions under which $\R$ can grow.

\section{Evolution of comoving curvature perturbations}

The metric  for scalar perturbations is given by
\begin{align}
    \dd s^2 &=-(1+2A)\dd t^2+2a\partial_iB \dd x^i\dd t+ a^2\left\{\delta_{ij}(1+2C)+2\partial_i\partial_jE\right\}\dd x^i\dd x^j  \, , \label{metricpert}
\end{align}
and the energy-momentum tensor for a minimally coupled scalar field is
\begin{align}
 T^0{}_0 &= -\left( \frac{1}{2}\dot{\phi}^2 + V(\phi) \right) -\dot{\phi} \dot{\delta\phi} + A \dot{\phi}^2 - V_{\phi} \delta \phi \,, \label{sempert00}  \\ 
 T^0{}_i &=   \partial_i \left( -\frac{\dot{\phi} \delta \phi}{a} \right) \,, \label{sempert0j}  \\ 
 T^i{}_j &=  \delta^{i}_{j} \left[ \frac{1}{2}\dot{\phi}^2 - V(\phi)+  \dot{\phi} \dot{\delta\phi}  - A \dot{\phi}^2 - V_{\phi} \delta \phi  \right] \, , \label{sempertij} 
\end{align}
where $\delta \phi$ is the perturbation of the scalar field.

The  curvature perturbation on comoving slices is a gauge invariant variable given by $C$ in the gauge in which $ T^0{}_i=0$, which for a single field takes the form  
\begin{align}
    \R= C + H \frac{\delta \phi}{\dot{\phi}} \, . \label{Rdef} 
\end{align}
The perturbed Einstein's equations in the comoving gauge are 
\begin{align}
 \frac{1}{a^2} \SPD [-\R+a H \sigma_c] &=-   \frac{\dot{\phi}^2 }{2 M_p^2} A_c \, , \label{eqzz}\\ 
A_c &= \frac{\dot{\R}}{H}  \, , \label{eqzi} \\
-\ddot{\R}-3H\dot{\R}+H \dot{A_c} +(2\dot{H}+3H^2)A_c  &=- \frac{ \dot{\phi}^2}{2 M_p^2}  A_c  \, , \label{eqii} \\
\dot{\sigma_c}+2H\sigma_c-\frac{A_c+\R}{a} &= 0 \, , \label{eqij}
\end{align}
where $\SPD \equiv \delta^{kl}\partial_k\partial_l$, $\sigma_c \equiv a \dot{E}_c -B_c$, and we denote with the supscript ${}_c$ quantities defined in the comoving slices gauge.
After appropriately manipulating the above equations \cite{Vallejo1:2018} it is possible to reduce the system of differential equations to a closed equation for $\R$, the Sasaki-Mukhanov's equation 
\begin{align}
    \ddot{\R}_k + 2 \frac{\dot{Z}}{Z} \dot{\R}_k + \frac{k^2}{a^2} \R_k &= 0 \, , \label{eqR}
\end{align}
where $Z\equiv \sqrt{\epsilon a^3}$ and ${\R}_k$  is the Fourier mode. 

The coefficients of this equation only depend on the background evolution through $a$ and $\epsilon$, so a priori no notion of entropy perturbation is required to solve them. It is nevertheless useful, as we will show later, to decompose $\delta P$ in an adiabatic and non-adiabatic component \cite{Wands:2000dp} to study the super-horizon behavior of $\R$. 
\section{Super-horizon growth of comoving curvature perturbations}
Eq.(\ref{eqR}) can be re-written in the form  
\begin{align}
    \frac{\dd}{\dd t}\left( Z^2 \dot{\R}_k \right) + a\epsilon k^2 \R_k &= 0 \, , \label{eqRM}
\end{align}
from which it is  possible, after re-expressing the time derivatives   in terms of derivatives with respect to the scale factor $a$,  to find a solution on super-horizon scales given by \cite{Romano:2016gop}
\begin{align}
  \R_k &\propto \int \frac{\dd a}{a} f \quad ; \quad f \equiv \frac{1}{HZ^2}= \frac{1}{Ha^3\epsilon} \, , \label{f}
\end{align}
corresponding to the constant mode $Z^2 \dot{\R}_k \propto const$.
From the above solution we can obtain the useful relations
\begin{align}
\dot{\R}_k &= \alpha f H \, , \label{Rdf} 
\end{align}
where $\alpha$ is an appropriate complex constant and we used the absolute value of $H$, because the same equation is also valid in a contracting Universe.
This mode in slow-roll models is sub-dominant, and is called decaying mode since the $1/a^3$ factor makes it rapidly decrease.
The above equation is the key to understand what is producing the super-horizon growth of comoving curvature perturbations. First of all it is clear that $f$ only depends on background quantities. This is already a first hint that entropy perturbations are not the cause of the growth of $\R_k$, which we will show explicitly in the last section.

From the definition of the slow roll parameter $\epsilon$ and the Friedman equations :
\bea
H^2&=&\frac{\rho}{3 M_p^2} \, , \label{eq:Friedmann} \\
\dot{H}&=& -\frac{\rho +  P}{2 M_p^2} \,,
\eea
we can get the useful relation
\be
\epsilon =\frac{3}{2}(1+w)=\frac{3}{2}\frac{\rho+P}{\rho} \, ,
\ee
which substituted in $f$
gives
\be
f = \frac{2}{3(1+w)Ha^3} \, ,
\ee
where $w\equiv P / \rho$.

From eq.(\ref{f}) we can immediately deduce that the general condition for super-horizon growth is  
\be
\frac{\dd f}{\dd a} \geq 0 \, , \label{gc}
\ee
or equivalently
\begin{equation}
    \frac{\dd f}{\dd a} =\frac{1}{\dot{a}} \frac{\dd f}{\dd t} = \frac{1}{a H} \dot{f} \geq 0 \, . 
\end{equation}
During inflation $aH>0$ and the general condition given in eq.(\ref{gc}) reduces to 
\begin{equation}
    \dot{f}\geq0 \, . \label{fdg}
\end{equation}
Note that in a contracting Universe $H<0$ and the condition for super-horizon growth  would be inverted, i.e. $\dot{f}<0$.

Eq.(\ref{fdg}) implies
\begin{align}
    \dot{f} &= \frac{\dd}{\dd t}\left(\frac{1}{HZ^2}\right)= -\frac{3-\epsilon+\eta}{Z^2}\geq0 \label{fdsr}\, ,
\end{align}
which gives the general condition for super-horizon growth in an expanding Universe, and when $Z^2>0$ implies $3-\epsilon+\eta\leq0$, in agreement with the results obtained in \cite{Ezquiaga:2018gbw}. Note that $Z^2>0$ is a natural assumption because models with $Z^2<0$ would at some point require a problematic phantom crossing \cite{Vikman:2004dc}.

The condition for super-horizon growth of $\R$ in terms of $w$ is
\be
\dot{f} = -\frac{1}{(1+w)^2 H a^3} \left[ \dot{w} + \frac{3H}{2}(1-w^2) \right] \geq 0 \, . \label{fdwd}
\ee

From the above equation we can  deduce some important conclusions  
\begin{itemize}
    \item in slow-roll inflation $w>-1$, $\dot{w} > 0$, and consequently $\dot{f}<0$, leading to the well known behavior of the decaying mode
    \item a sufficiently fast decrease of $w(t)$, i.e. $\dot{w}< -\frac{3H}{2}(1-w^2)$, can give $\dot{f}>0$, and consequently induce the super-horizon evolution of the "decaying" mode  
\end{itemize}
Note that in  drawing the conclusions above we have assumed again $w>-1$, to avoid phantom crossing.
As we will see later the second case is the mechanism which causes the growth of $\R$ in quasi-inflection inflation, a model in which there is first an increase in $w$ and then a sudden decrease, during which the "decaying" modes grows.

\section{Super-horizon growth of $\R$ in globally adiabatic model}
Globally adiabatic (GA) models were investigated in details \cite{Romano:2016gop}, including different examples such as generalized ultra slow roll inflation, and Lambert inflation. It is interesting to understand what causes the super-horizon growth of $\R$ in these models in the light of eq.(\ref{fdwd}).

From the time derivative of $w$ and the continuity equation 
\be
\dot{w} = \frac{\dot{\rho}}{\rho} \left(c_w^2 - w \right) \quad , \quad \dot{\rho} + 3 H (1+w)\rho=0 \, , 
\ee 
where $c_w^2\equiv \dot{P}/\dot{\rho}$, we can also re-write the condition for super-horizon growth as
\be
\dot{f} =  \frac{2 c_w^2 - w -1}{(1+w)a^3}
\geq 0 \, , \label{fdcw}
\ee
which assuming $w>-1$ implies
\be
c^2_w\geq \frac{w+1}{2} \,.
\ee
The same kind of results can be easily generalized to the class of GA models studied in  \cite{Romano:2016gop} corresponding to $\epsilon \propto a^{-n}$, $c^2_w \approx \frac{n-3}{3}, w\approx -1$, so that $\dot{f}>0$ implies $n>3$, which is indeed in agreement with the result obtained in \cite{Romano:2016gop}. Note that this is just an alternative way to interpret the general condition given in eq.(\ref{fdwd}) in terms of the adiabatic speed $c^2_w$, but the super-horizon growth of $\R$ could be understood equivalently as the consequence of a fast decrease of $w=2\epsilon/3-1 = 2 a^{-n}/3 -1$.

\section{A general relation between entropy
perturbations and comoving curvature pertubations}

The standard definition of  non-adiabatic pressure for a  single field is  \cite{Wands:2000dp} 
\begin{align}
    \delta P_{na} \equiv \delta P - c_w^2 \delta \rho   \, , \label{deltaPnad} 
\end{align}
where $c^2_w=\dot{P}/\dot{\rho}$. It is easy to verify that $\delta P_{na}$ is gauge invariant and corresponds to the pressure perturbation in the uniform density gauge in which $\delta\rho=0$, i.e. $\delta P_{na}=\delta P_{ud}=\delta P_{\delta\rho=0}$, where the subscript ${}_{ud}$ stands for uniform density. This is indeed the gauge in which the relation between curvature perturbations on uniform density slices $\zeta$ and adiabaticity were originally studied \cite{Wands:2000dp}. 

In the comoving slices gauge, defined by the condition $\delta \phi_c =0$, from eq.(\ref{sempert00}) and eq.(\ref{sempertij}) we have
\begin{align}
    \delta \rho_c &= \delta P_c = - \dot{\phi}^2 A_c \, .
\end{align}
Using its gauge invariance, $\delta P_{na}$  can be computed  in the comoving slices gauge as
\begin{align}
    \delta P_{na} &=\delta P_c - c_w^2 \delta \rho_c  =\left(c_w^2 -1 \right) \dot{\phi}^2 A_c=\left(c_w^2 -1 \right) \frac{\dot{\phi}^2}{H} \dot{\R} \, . \label{deltaPnadssf} 
\end{align}
where in the last equality we have used the perturbed Einsteins's equation $\dot{\R}=H A_c$.

The above expression can also be written in the form
\begin{align}
    \delta P_{na} &= \frac{2}{3}\frac{\epsilon \rho}{H}  \left(c_w^2 -1 \right) \dot{\R} = \frac{ \rho}{H} \left(w +1 \right)  \left(c_w^2 -1 \right) \dot{\R}  \, , \label{deltaPnadw}
\end{align}
 from which it can be seen  that $\dot{\R}$ could be arbitrary large, but as long as $c_w^2 \approx 1$  the non-adiabatic pressure perturbation $\delta P_{na}$ can  be arbitrary small. 
 The limiting case in which $c_w^2=c_s^2=1$ corresponds to a globally adiabatic system  \cite{Romano:2016gop,Romano:2015vxz}, that  is adiabatic on any scale, i.e. $\delta P_{na}=0$, but for which $\R$ grows on super-horizon scales.

In order to  understand the relation between $\R$ and  $\delta P_{na}$ it is useful to re-write eq.(\ref{deltaPnadw}) as 
\be
\dot{\R}=C_f \delta P_{na} \, , \label{def:Cf} 
\ee
where $C_f$ is the conversion factor between non adiabatic and  curvature perturbations, given by
\be
C_f = \frac{1}{3 M_p^2 H (1+w)(c_w^2-1) } \, .
\label{Cfbd} 
\ee

Contrary to previous results  derived in the uniform density gauge assuming the gradient term can be neglected on super-horizon scales \cite{Wands:2000dp}, the above relation is more general because is valid on \textit{any scale} for any minimally coupled single scalar field model, also on sub-horizon scales where gradient terms are large, and does not rely on any assumption about the relation between the curvature on uniform density slices $\zeta$ and $\R$.
Eq.(\ref{def:Cf}) is also more general than eq.(\ref{Rdf}), which is also only valid on super-horizon scale.

We will show in the next section that the behavior of the conversion factor is the main cause of the super-horizon growth of $\R$, not the entropy perturbations in themselves.
It is also useful to derive  this relation for later use
\bea
|\dot{\R}|&=&|C_f| |\delta P_{na}| \, . \label{def:Cfabs} 
\eea

Note that in the case of globally adiabatic systems the situation is more complex because the equality $c^2_s=c^2_w$ causes a divergence of the conversion factor, while at the same time $\delta P_{na}=0$. In this case non adiabatic pressure perturbations and comoving curvature perturbations are completely independent and only an explicit calculation of $\dot{f}$ allows to find  the super-horizon behavior of $\R$ \cite{Romano:2016gop}.

\section{Application to quasi-inflection inflation}
\begin{figure}
\centering
\includegraphics[width=.8\textwidth]{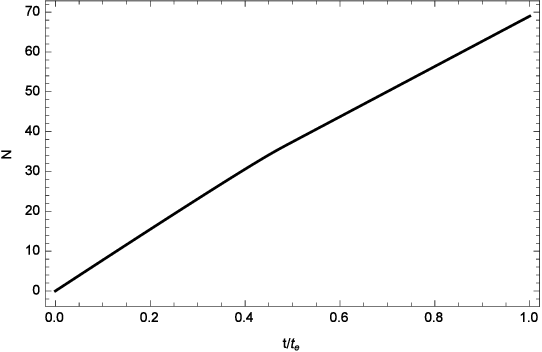}
\caption{The number of e-folds before the end of inflation $N$ is plotted as a function of $t/t_e$, where $t_e$  is the inflation end time, defined by the condition $\epsilon=1$.}
\label{NE}
\end{figure}

\begin{figure}
\centering
\includegraphics[width=0.8\textwidth]{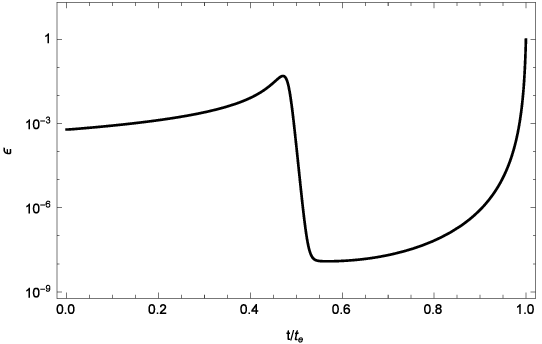}
\caption{The slow-roll parameter $\epsilon$ is plotted as a function of the e-folds number.}
\label{epsilont}
\end{figure}

\begin{figure}
\centering
\includegraphics[width=0.8\textwidth]{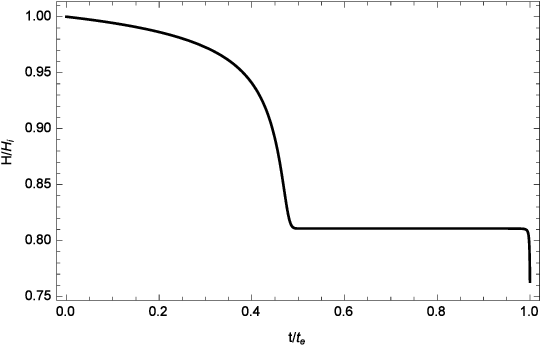}
\caption{The Hubble parameter is plotted as a function of time.}
\label{Ht}
\end{figure}

As an application of the general results obtained in the previous sections here we study the case of quasi-inflection inflation, a single scalar field model minimally coupled to gravity with  potential 
\cite{Garcia-Bellido:2017mdw,Ezquiaga:2018gbw,Germani:2017bcs} 
\begin{align}
   V(\phi) &=  \frac{\alpha}{12}\frac{6 v^2\phi^2-4 v \phi^3 + 3  \phi^4}{(1+\beta \phi^2)^2} \, . \label{V}
\end{align}
The plots given in the paper are obtained by solving numerically the system of coupled differential equations (\ref{eqH}) and (\ref{eqphi}) with the following choice of parameters 
\begin{align}
    \alpha &= 2.97 \times 10^{-7} \quad , \quad v= \sqrt{.108} M_p  \quad , \quad \beta = \frac{1}{v^2} \left\{\frac{1}{3} \left[ 2+\left( \frac{7}{2} \right)^{2/3} \right] - 10^{-4} \right\} \, .
\end{align}
In figs.(\ref{NE}-\ref{Ht}) we plot the number of e-folds from the beginning of inflation $N$, the slow-roll parameter $\epsilon$ and the Hubble parameter $H$ as functions of time in units of the inflation end time $t_e=6.99 \times 10^{6} M_p^{-1}$, defined as the time when $\epsilon=1$. We choose the initial conditions in order to satisfy the Planck constraints according to $\{\phi_i=10.75 v, \dot{\phi}_i=-3.91 \times 10^{-7} M_p^2\}$, and the corresponding initial value of the Hubble parameter is $H_i=1.11 \times 10^{-5}M_p$. 
As shown in fig.(\ref{epsilont}) there is a region where $\epsilon$ is quickly decreasing, and we will show in the next section that this is indeed the cause of the super-horizon growth of curvature perturbations.  

\section{Behavior of perturbations}
\begin{figure}[H]
\centering
\includegraphics[width=\textwidth]{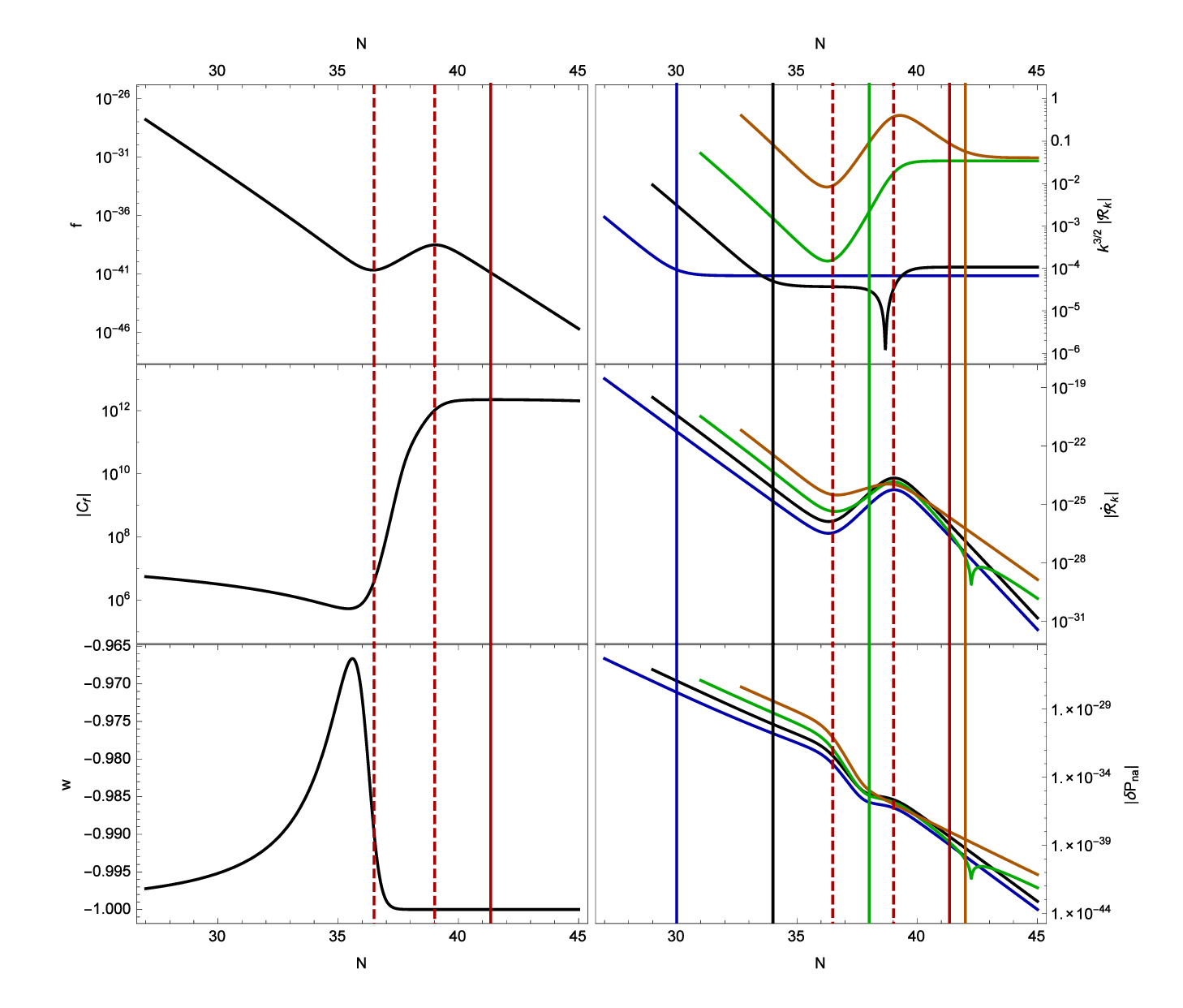}
\caption{On the left the background quantities $f,|C_f|,w$ are plotted as functions of e-folds number. On the right the norm of the curvature perturbation, its time derivative and non adiabatic pressure perturbation are plotted in the same e-folds interval. The dashed lines delimit, the e-folds interval $(N_1,N_2)$ during which the function $\dot{f}\geq0$. Denoting as $\tau_1$ the conformal time at $N_1$ and the comoving scale leaving the horizon at that time as $k_1=-1/\tau_1$, different modes of perturbations are plotted in blue ($k \approx 1.75 \times 10^{-3} k_1$), black ($k \approx 8.85 \times 10^{-2} k_1$), green ($k \approx 4.53  k_1$) and orange ($k \approx 2.47 \times 10^{2} k_1$). The vertical solid red line corresponds to the e-folds number $N_i$ at which the field reaches the quasi-inflection point, and the other vertical solid lines denote the horizon crossing e-folds number of the corresponding modes. The super-horizon evolution of curvature perturbation  occurs in the e-folds  interval in which $\dot{f}\geq0$,  or equivalently $\dot{w}<0$, or $|C_f|$ has a large variation.} 
\label{gp}
\end{figure}

\begin{figure}[H]
\centering
\includegraphics[width=\textwidth]{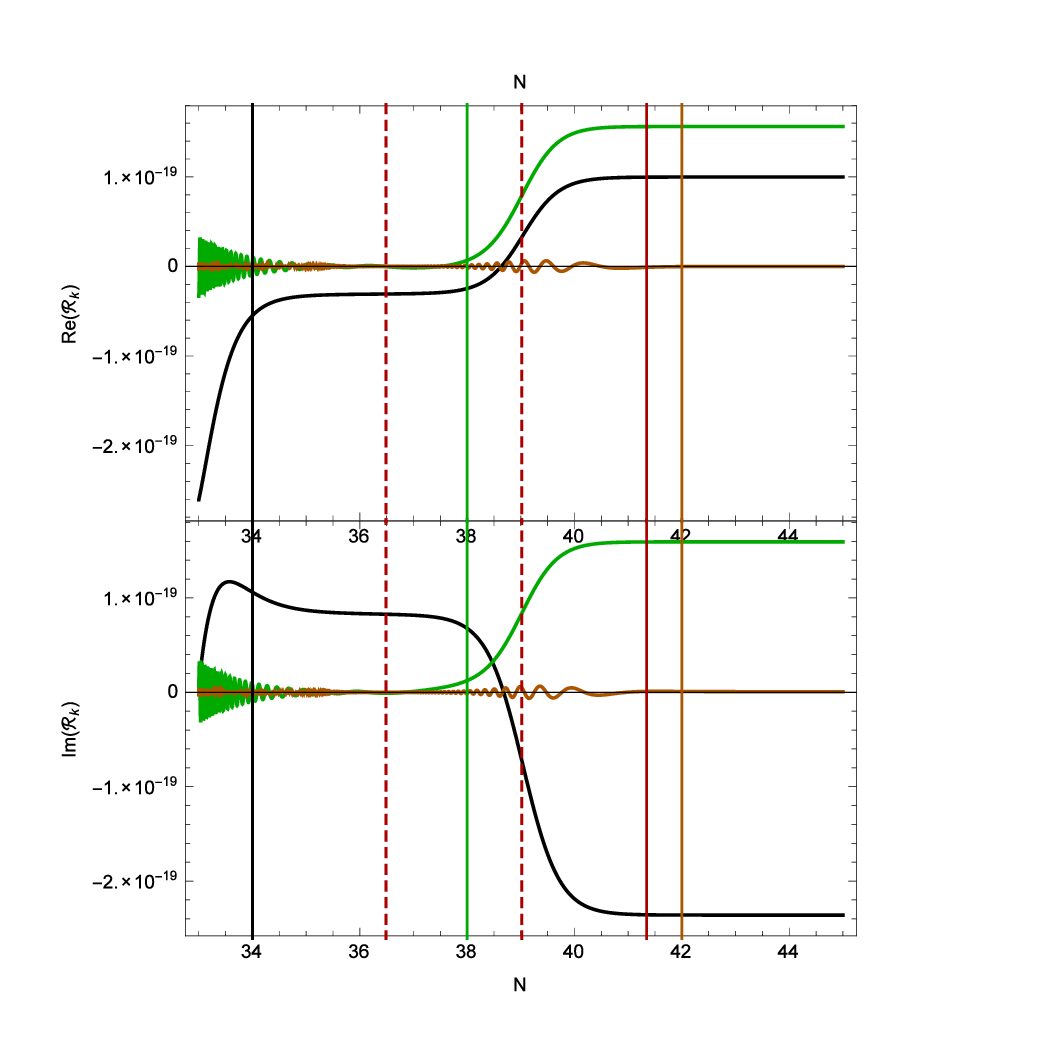}
\caption{The real and imaginary part comoving curvature perturbations are plotted as functions of e-folds number for the same comoving scales of fig.(\ref{gp}). After horizon crossing both the real and imaginary parts of $\R_k$ are monotonous functions of time, in agreement with eq.(\ref{Rdf}). The vertical solid black (green and orange) line(s) correspond(s) to the horizon crossing e-folds number, the vertical solid red line corresponds to the e-folds number at which the field reaches the quasi-inflection point and the dashed lines delimit the same e-folds interval $(N_1,N_2)$ defined for  fig.(\ref{gp}).}
\label{RkEp}
\end{figure}

We solve numerically the equation for curvature perturbations  imposing initial conditions corresponding to the Bunch-Davies vacuum  when the modes are deep inside the horizon. The curvature evolution is shown in figs.(\ref{gp}) and (\ref{RkEp}), from which it can be seen that the mode is not freezing after horizon crossing.

As can be seen in fig.(\ref{gp}) the super-horizon evolution of $|\R_k|$ occurs during the time interval in which $f$ is a growing function of time, whose limits are denoted in this and all other figures with dashed vertical red lines, in agreement with the general condition obtained in the previous section. 
During the same interval $|\dot{\R}_k|$ is growing, due to the growth of the norm of the conversion factor $|C_f|$, in agreement with eq.(\ref{def:Cfabs}), despite $|\delta P_{na}|$ is decreasing in the same interval, clearly showing that the behavior of the conversion factor is the main cause of the evolution of curvature perturbations, not entropy perturbations, contrary to what claimed in \cite{Ezquiaga:2018gbw}.

The key quantities to understand the evolution of different modes are the co-moving scales $k_1=-1/\tau_1$ and $k_2=-1/\tau_2$ corresponding to the modes leaving the horizon at the conformal times $\tau_1,\tau_2$ when the condition $\dot{f}>0$ respectively starts and ends to be satisfied. Larger scale modes ($k<<k_1$) exiting the horizon much earlier than $\tau_1$  are not affected (see the blue line), while modes leaving the horizon around the time interval ($\tau_1$,$\tau_2$) are affected, as shown in the case of the black, green and orange lines. Smaller scale modes ($k>>k_2$) exiting the horizon much after $\tau_2$ are deeply inside the horizon between $\tau_1$ and $\tau_2$, and for this reason are not affected since the gradient term dominates.
Note that the curvature growth happens before the inflection point, showing that the motivation for considering potentials with an inflection point based on the slow-roll formula $P^{1/2}_{\R} \approx 1/V'$ is miss-leading, as pointed out in \cite{Germani:2017bcs,Motohashi:2017kbs}, since the curvature growth takes place during a regime of strong slow-roll violation, which can happen also without an inflection point, in presence of other types of features of the potential \cite{Ozsoy:2018flq}. The violation of the slow-roll  conditions could also have other important effects such as the violation of approximate consistency conditions \cite{Chung:2005hn} based on the assumption of slow-roll.

\begin{figure}[H]
\centering
\includegraphics[width=\textwidth]{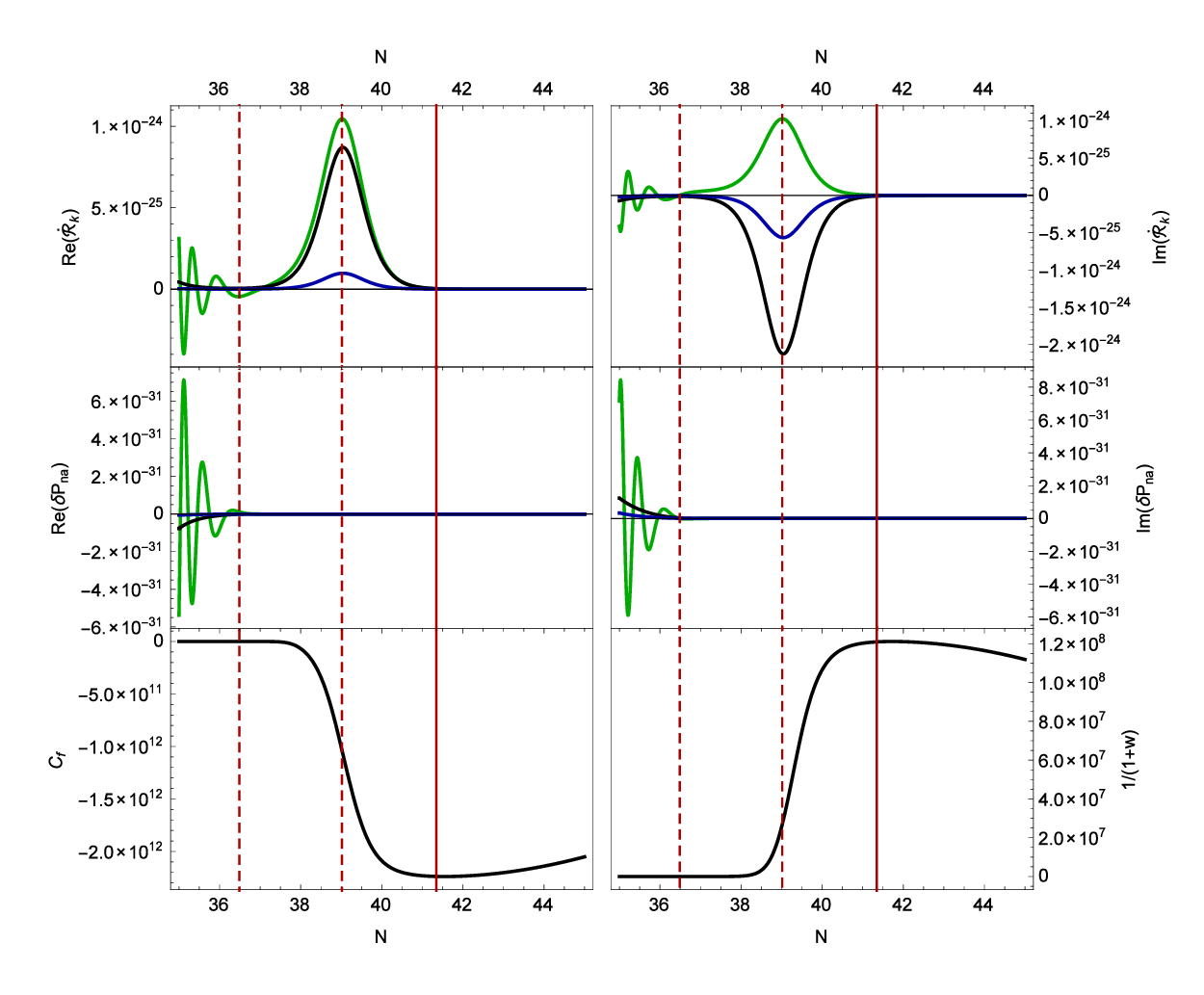}
\caption{The real and imaginary parts of $\dot{\R}_k$ and $\delta P_{na}$, the conversion factor $C_f$ and the quantity $(1+w)^{-1}$ are plotted as functions of e-folds number. We plot two modes which have exited the horizon before interval $N_1$, the blue ($k \approx 1.75 \times 10^{-3} k_1$), and black ($k \approx 8.85 \times 10^{-2} k_1$) lines, and one mode which is leaving the horizon during the $(N_1,N_2)$ interval, the  green line($k \approx 4.53  k_1$).  The large increase of the amplitude of the oscillations of $\dot{\R}_k$ is caused by the conversion factor $C_f$, which is mainly due to the sudden change of $w(t)$. The amplitude of the oscillations of the entropy perturbation $\delta P_{na}$ is  decreasing, and as such cannot be considered the cause of the super-horizon increase of  $\R_k$. The vertical solid red line corresponds to the e-folds number at which the field reaches the quasi-inflection point and the dashed lines delimit the same e-folds interval $(N_1,N_2)$ defined for  fig.(\ref{gp}).}
\label{GP2}
\end{figure}

Fig.(\ref{RkEp}) and fig.(\ref{GP2}) show that the real and imaginary parts of $\R_k$ are both monotonous functions of time, as expected from eq.(\ref{Rdf}). Note that in the same time interval $|\R_k|$ is not  monotonous because the norm  mixes the imaginary and real parts.   

\begin{figure}[H]
\centering
\includegraphics[width=.9\textwidth]{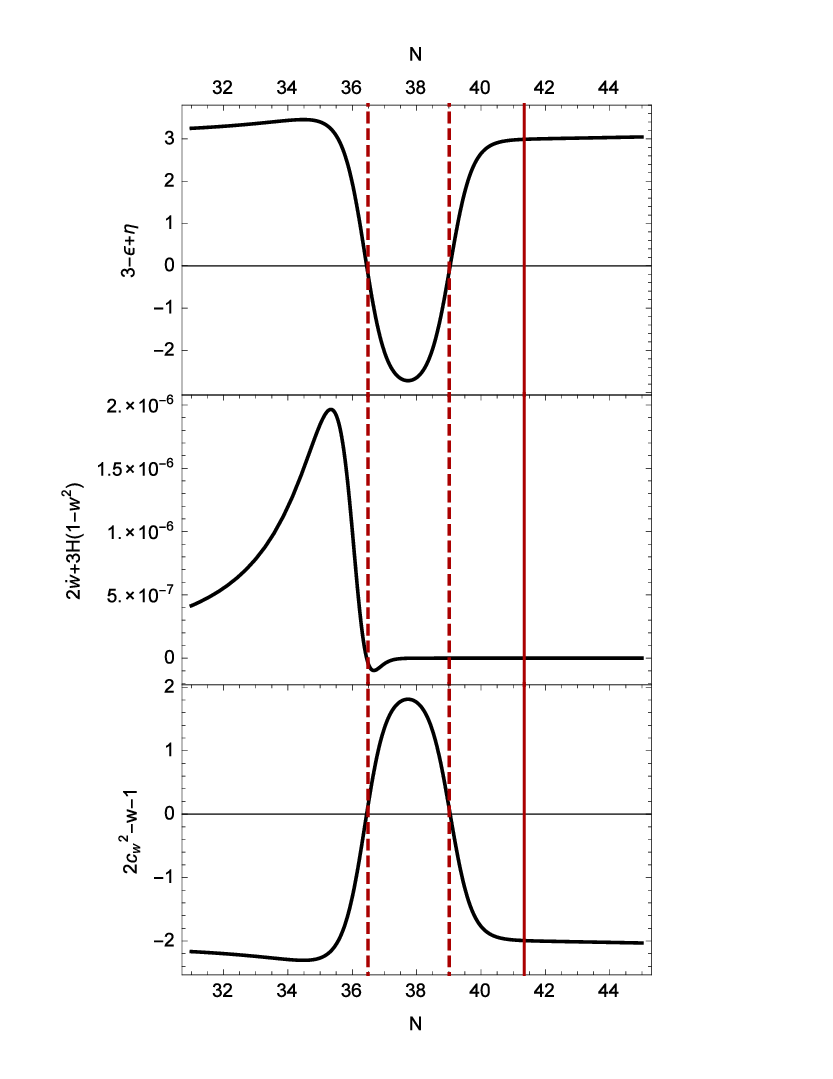}
\caption{The background quantities $3-\epsilon+\eta$, $2\dot{w}+3H(1-w^2)$ and $2c_w^2-w-1$ are plotted as functions of e-folds number.
The super-horizon evolution of curvature perturbations occurs when $\dot{f}\geq0$, which according to eqs.(\ref{fdsr}-\ref{fdwd}) and eq.(\ref{fdcw}) implies the three equivalent conditions $3-\epsilon+\eta\leq0$, $2\dot{w}+3H(1-w^2)\leq0$ or $2c_w^2-w-1\geq0$. The vertical red solid and dashed lines are plotted to denote the same e-folds numbers indicated in the same way in  fig.(\ref{gp}).} 
\label{fd}
\end{figure}

In fig.(\ref{fd}) are shown the time intervals during which the conditions $3-\epsilon+\eta\leq0$, $2\dot{w}+3H(1-w^2)\leq0$ and $2c_w^2-w-1\geq0$ are satisfied, which according to eqs.(\ref{fdsr}-\ref{fdwd}) and eq.(\ref{fdcw}) imply $\dot{f}\geq0$. The different plots allow to understand in alternative ways the origin of the super-horizon evolution of $\R_k$ in terms of the behavior of different background quantities such as the adiabatic sound speed $c_w$, the equation of state $w(t)$ or the slow roll parameters $\epsilon$ and $\eta.$

\begin{figure}[H]
\centering
\includegraphics[width=\textwidth]{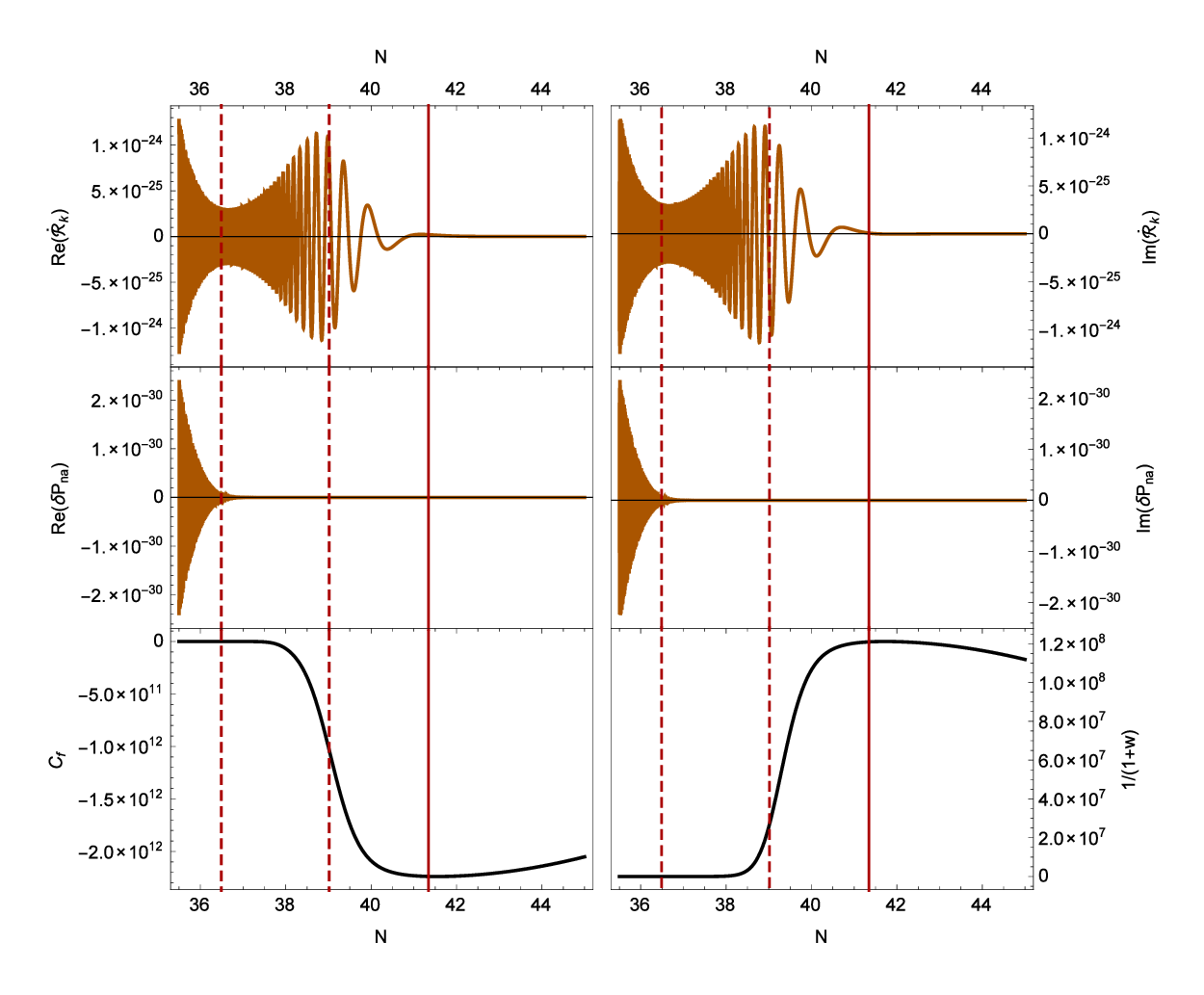}
\caption{The rapidly oscillating real and imaginary parts of $\dot{\R}_k$ and $\delta P_{na}$, the conversion factor $C_f$ and the quantity $(1+w)^{-1}$ are plotted as a function of the e-folds number. The perturbations corresponds to a mode which is sub-horizon ($k \approx 2.47 \times 10^{2} k_1$) in the interval $(N_1,N_2)$ . The large increase of the amplitude of the oscillations of $\dot{\R}_k$ is caused by the conversion factor $C_f$, which is mainly due to the sudden change of $w(t)$. The amplitude of the oscillations of the entropy perturbation $\delta P_{na}$ is  decreasing, and as such cannot be considered the cause of the sub-horizon increase of  $\R_k$. The vertical solid red line corresponds to the e-folds number at which the field reaches the quasi-inflection point and the dashed lines delimit the same e-folds interval $(N_1,N_2)$ defined for fig.(\ref{gp}).}
\label{GP32}
\end{figure}

Since eq.(\ref{def:Cf}) is valid on any scale, it can also be used to understand the evolution of curvature perturbations for modes which are sub-horizon during  the time interval $(\tau_1$,$\tau_2)$. As shown in fig.(\ref{GP32}) the conversion factor is in fact causing a sub-horizon enhancement of curvature perturbations, despite entropy perturbations are decreasing.

\section{Conclusions}
The super-horizon growth of comoving curvature perturbations can have very important observable effects, such as for example the production of primordial black holes.
We have investigated what are the general causes of this phenomenon in single scalar field models and found that non adiabatic perturbations are not the main cause, but it is rather the evolution of the background which determines this growth.
We have shown that the key quantity to consider is the equation of state $w$, and that in an expanding Universe with $w>-1$ the super-horizon growth is due to a sufficiently fast decrease of $w$.
We have also given an equivalent condition for the super-horizon evolution in terms of the adiabatic sound speed, $c_w^2\geq (w+1)/2$.

We have then derived a  general  relation between  the time derivative of comoving curvature perturbations and entropy perturbations, in terms of a conversion factor depending on the background evolution. Contrary to previous results  derived in the uniform density gauge assuming the gradient term can be neglected on super-horizon scales, the relation is  valid on any scale for any minimally coupled single scalar field model, also on sub-horizon scales where gradient terms are large.
The case of globally adiabatic systems is peculiar because entropy perturbations are vanishing on any scale and the growth of curvature perturbations cannot be consequently understood in terms of the conversion factor, but only in terms of the behavior of the adiabatic sound speed or equivalently of the function $f$.

Applying the general relation between comoving curvature perturbations and entropy perturbations to  the case of quasi-inflection inflation, we found that the growth of $\R$ is due to a sudden and large variation of the conversion factor between non adiabatic and curvature perturbations.
The super-horizon growth of $\R$ occurs during a time interval in which entropy perturbations decrease, clearly showing that entropy perturbations are not the main cause of the growth of $\R$.

In the future it will be interesting to consider other  models which could produce PBH due to the super-horizon growth of $\R$, focusing the search on those which can give a sufficiently fast decrease of the equation of state $w$. For example single field models with local features of the potential could be good candidates \cite{GallegoCadavid:2016wcz,Cadavid:2015iya} or other types of features \cite{Ozsoy:2018flq,Hazra:2010ve,Arroja:2011yu,Chen:2006xjb,Joy:2007na}.
It could also be interesting to study other effects of  the violation of the slow-roll regime,  such as the violation of approximate consistency conditions \cite{Chung:2005hn} based on the assumption of slow-roll.
\acknowledgments
We thank Juan Garc\'ia-Bellido and Jose Mar\'ia Ezquiaga for useful comments and discussions. 
This work was supported by the UDEA Dedicacion exclusiva and
Sostenibilidad programs and the CODI projects 2015-4044 and 2016-10945.

\bibliographystyle{h-physrev4}
\bibliography{mybib}
\end{document}